\documentclass[pre,preprint,english]{revtex4}
\usepackage{fancyhdr}
\usepackage{amsmath,amsfonts,amssymb}
\usepackage{graphicx}
\usepackage{color}
\usepackage{bbm}
\newcommand{\eref}[1]{(\ref{#1})}
\newcommand{\tr}{\operatorname{tr}}

\begin{document}
\title{A sub-determinant approach for pseudo-orbit expansions of spectral determinants in quantum maps and quantum graphs
}
\author{Daniel Waltner}
\address{Weizmann Institute of Science, Physics Department, Rehovot, Israel,\\ Institut f\"ur Theoretische Physik, Universit\"at Regensburg, D-93040 Regensburg, Germany}
\author{Sven  Gnutzmann}
\address{School of Mathematical
  Sciences, University of Nottingham, Nottingham NG7 2RD, UK}
\author{Gregor Tanner}
\address{School of Mathematical
  Sciences, University of Nottingham, Nottingham NG7 2RD, UK}
\author{Klaus Richter}
\address{Institut f\"ur Theoretische Physik, Universit\"at Regensburg, D-93040 Regensburg, Germany}


\begin{abstract}
We study implications of unitarity for pseudo-orbit expansions of the spectral determinants of
quantum maps and quantum graphs. In particular, we advocate to group pseudo-orbits into sub-determinants. We
show explicitly that the cancellation of long orbits is elegantly described
on this level and that unitarity can be built in using a simple
sub-determinant identity which has a non-trivial interpretation in terms
of pseudo-orbits. This identity yields much more detailed relations between pseudo orbits of different length than known previously.
We reformulate Newton identities and the spectral density in terms of
sub-determinant expansions and point out the implications of the sub-determinant
identity for these expressions. We analyse furthermore the effect of the identity on
spectral correlation functions such as the auto-correlation and parametric cross
correlation functions of the spectral determinant and the spectral form factor.
\end{abstract}
\maketitle

\section{Introduction}

\subsection{Overview}

When calculating quantum spectra with the help of periodic-orbit sums such as, for
example, arising from semiclassical expressions, one typically encounters problems due to divergencies resulting from summing over a large number of periodic orbits which grows exponentially with length. This is in particular the case for quantum systems whose underlying classical dynamics is chaotic \cite{Eck89}.  To apply these periodic-orbit expressions for determining quantum spectra, the number of relevant orbits needs to be reduced.
This is either achieved by reordering the orbit contributions making use of
cancellations such as is done in the cycle expansion \cite{Artu} or one can also utilise unitarity of the quantum dynamics leading to additional relations between
the coefficients of the  characteristic polynomial and thus to finite sums over pseudo-orbits \cite{Berry,Tan91,Georgeot,Bog92,Bogomolny,BHJ12,Texier}.

A related problem is the semiclassical calculation of spectral correlation functions.
They are conjectured to follow Random Matrix Theory (RMT) for quantum systems with chaotic classical
limit. Establishing this connection explicitly using semiclassical periodic-orbit formulae for
the spectral form factor could only be achieved fairly recently following the work in \cite{Sieber}.
This calculation has been extended in \cite{Heusler} yielding the full
spectral form factor as predicted by RMT for times smaller than the Heisenberg time $T_H$.
(This is the time needed to resolve distances of the order of the mean level spacing in the Fourier-transformed spectrum).
The spectral form factor for times larger than $T_H$ has been obtained using semiclassical periodic-orbit
expressions in \cite{HeuslerI}. The calculation is based on a generating function approach containing two
spectral determinants both in the numerator and denominator at four different energies. The derivation makes
explicit use of the fact that the spectral determinant is real for real energies. Although this is obvious
from its definition, Eq.\ (\ref{eq5}) below, it is not clear a-priory when considering the representation
of the spectral determinant containing periodic-orbit sums.
A real spectral determinant in terms of periodic orbits can only be
semiclassically obtained by exploiting
periodic-orbit correlations due to unitarity.\\

The above problem illustrates that we need a better understanding of correlations between periodic
orbits and in particular correlations between long and short orbits.
To analyse these correlations in more detail, we study here
quantum unitary dynamics described in terms of finite dimensional unitary matrices,  i.e.\ quantum unitary maps. In this case periodic orbits refer to products of elements of the describing unitary matrix with their indices forming a closed cycle. We give later an interpretation in terms of periodic orbits on quantum graphs \cite{Kottos} where exact periodic-orbit expansions for spectral quantities exist. These expansions are of similar form as the semiclassical approximations obtained for more general systems.  We will in particular advocate to consider spectral quantities in terms of sub-determinant expansions.  By this we obtain a much more detailed relation between contributions from orbits of different length for closed systems. The previously known relations \cite{Bog92} only connect the {\it overall} (summated) contributions to spectral quantities from long and short orbits. We, however, derive a relation between contributions from short orbits within {\it different} parts of the system and its corresponding complementary orbits. Such an identity is of particular importance when spatially inhomogeneous effects, such as a magnetic field, that affect the contributions from {\it different} orbits of the {\it same} length differently are considered. We expect it also to lead to simplifications in the diagrammatic expansions in \cite{Texier}.
We afterwards will derive sub-determinant expressions for a range of important spectral quantities and consider these for
examples such as quantum graphs.

The paper is structured as follows: We first introduce the spectral determinant and explain the known implications of
unitarity for this quantity. We analyse in Sec.\ \ref{sec:po-expansion} further implications of
unitarity on pseudo-orbit expansions. In this context, we present a sub-determinant identity for unitary matrices  and explain how it yields the considered relation between short and long orbit contributions. We discuss the implications of this identity on Newton identities and a pseudo-orbit expansion of the spectral density.
In Sec.\ \ref{sec:applications}, we derive expressions for spectral correlation functions such as the
auto-correlation and the parametric cross correlation function of the spectral determinant and the spectral
form factor in terms of sub-determinant expansions.  Implications due to the sub-determinant identity will be discussed.

\subsection{Some basic properties of the characteristic
  polynomial of a matrix}

Consider a general complex matrix $U$ of dimension $N$.
Its characteristic polynomial is given by
\begin{equation}
  P_U(z)\equiv
  \det \left(z - U\right)
  =
  \sum_{n=0}^N (-1)^{N-n} a_{N-n} z^n
  =
  \prod_{n=1}^N (z-z_n),
  \label{char_poly}
\end{equation}
where the complex numbers $z_n$ are the eigenvalues
of $U$. The complex coefficients $a_n$ of the polynomial in (\ref{char_poly})
will be at the centre of
interest in this article. Here, $a_0=1$ and the remaining $N$ coefficients $a_n$, $n=1,\dots,N$,
are $N$ complex numbers which contain the same information as the $N$
eigenvalues $z_n$. Note that the characteristic polynomial is
invariant under conjugation  $U \mapsto C U C^{-1}$ with a
non-singular
matrix $C$. The coefficients $a_n$ are thus matrix invariants (as are
the eigenvalues) and can be expressed in terms of other matrix
invariants such as traces of powers of $U$.
Indeed, expressions for the coefficients $a_n$ in Eq.\ (\ref{char_poly}) in
terms of eigenvalues or  traces of $U$ can be easily written down, for instance
\begin{eqnarray*}
a_{1}&=&\sum_{n=1}^N z_n =  \mathrm{tr}\, U,\\
a_{2}&=&\frac{1}{2} \sum_{n \neq m} z_n z_m=
\frac{1}{2}\left( \mathrm{tr}^2\ U  -\mathrm{tr}\ U^2 \right).
\end{eqnarray*}
Similar formulae expressing the $a_n$'s in terms of traces hold for all $n$ \cite{fritz_qsc}.
Note however, that $a_N= \prod_{n=1}^N z_n = \det U$ has a much
simpler expression in terms of the determinant of $U$.

Alternatively one may express the coefficients in terms of
sub-determinants of $U$. Denote the set $\mathcal{I}=\{1,2,\dots,N\}$
and let $\Gamma \subset \mathcal{I}$ be some subset of $\cal I$ of cardinality $|\Gamma|$. Note that there are $2^N-1$ nontrival subsets of $\mathcal{I}$ that we write in the form $\left\{\Gamma\right\}_{j=1}^{2^N-1}$.  Then $\Gamma$ defines a quadratic  $|\Gamma|\times |\Gamma|$ submatrix
$U_\Gamma$   which is obtained
from $U$ by keeping only those rows and columns with indices
belonging to $\Gamma$. We will denote the determinant of $U_\Gamma$ as
\begin{equation}
  d_\Gamma=\det U_\Gamma \, .
  \label{subdet}
\end{equation}
Using the linearity properties of the determinant
with respect to its rows (or columns), it is then straight forward to show that
\begin{equation}
  a_n=\sum_{\Gamma \subset \mathcal{I}:\ |\Gamma|=n }
  d_\Gamma \ .
  \label{coeff_Gamma}
\end{equation}
The sum extends over the ${N \choose n}$ different choices of $n$ rows
(and the corresponding columns) that build the sub-matrix $U_\Gamma$.
While $a_n$ is a matrix invariant it is noteworthy that
this is in general not the case for the individual
contributions
$\det U_\Gamma$.

\section{On
  pseudo-orbit expansions in terms of determinants}
\label{sec:po-expansion}

\subsection{Basic relations}

Let us now consider the characteristic polynomial and some related
spectral functions for the specific case of {\em unitary} matrices  $U$.
We will keep the discussion general here and will only later refer to $U$ as the evolution matrix for a quantum system.

A unitary matrix $U$ of dimension $N $ has $N$
uni-modular eigenvalues $z_n= e^{i
  \theta_n}$.
This implies the functional equation
\begin{equation}
  P_U(z)=(-z)^N e^{i \phi} P_U(1/z^*)^*
  \label{func_eq}
\end{equation}
for its characteristic polynomial
where $z^*$ denotes the complex conjugate
of $z$ and $e^{i \phi} =\det\ U =  a_N$. Comparing coefficients of $z^n$ on
both sides of the functional equation (\ref{func_eq}) results in the
explicit
relation
\begin{equation}
  a_{N-n}=e^{i \phi} a_n^*
  \label{coeff_identity}
\end{equation}
between the coefficients of the characteristic polynomial.
Below in Sec.\ \ref{sec:sub_det}, we will generalise this relation to individual
determinants of sub-matrices contained in the
coefficient $a_n$ according to (\ref{coeff_Gamma}).

For unitary maps it is useful to introduce the following variant
of the characteristic polynomial, the so called zeta function,
\begin{equation}
  \zeta_U(\theta)=
  e^{-i N \theta}P_{U}(e^{i\theta})=
  \det \left(\mathbbm{I}- e^{- i \theta} U \right)
  = \sum_{n=0}^N (-1)^n a_n e^{-i\theta n}\ .
  \label{zeta_U}
\end{equation}
This is a $2\pi$ periodic function in the variable
$\theta$ which vanishes exactly at the spectrum of real
eigenphases $\{ \theta_n\}_{n=1}^N$.
In terms of sub-determinants (\ref{subdet}) one may also write
\begin{equation}
  \zeta_U(\theta)=\sum_{\Gamma}
  d_\Gamma e^{-i (\theta+\pi) |\Gamma|}
  \label{zeta_Ud}
\end{equation}
where the sum is over all subsets $\Gamma\subset \mathcal{I}$
including the empty set $\Gamma=\emptyset$ with $|\Gamma|=|\emptyset|=0$ for
which we set $d_{\emptyset}=1$.

The functional equation (\ref{func_eq}) implies that
\begin{equation}
\label{eq5}
  Z_U(\theta)= e^{iN\frac{\theta+\pi}{2}-i \frac{\phi}{2}}\zeta_U(\theta),
\end{equation}
usually referred to as spectral determinant,
is real for real $\theta$, i.e. $Z_U(\theta)^*=Z_U(\theta)$.

Another spectral function which will be discussed later is the
density of states
\begin{equation}
  \rho(\theta) = \sum_{n=1}^N \delta_{2\pi}(\theta -\theta_n)
\end{equation}
where $\delta_{2\pi}(x)\equiv \sum_{n=-\infty}^\infty \delta(x+2\pi
n)$
is the $2\pi$ periodic $\delta$-comb.
The density of states can be expressed as
\begin{equation}
  \rho(\theta) = \frac{1}{\pi} \frac{d}{d \theta}
  \mathrm{Im} \log Z_U(\theta- i \epsilon)
  = \frac{N}{2\pi} + \frac{1}{\pi} \frac{d}{d \theta}
  \mathrm{Im} \log \zeta_U(\theta- i \epsilon)\
  \label{dos_zeta}
\end{equation}
in the limit $\epsilon \to 0$.
This expression directly leads to the trace formula which expresses
the density of states in terms of periodic orbits. We will discuss
this in Sec.\ \ref{sec:pseudo} together with a novel expansion in terms
of sub-determinants presented in the next section.

\subsection{A sub-determinant identity for unitary matrices}
\label{sec:sub_det}

We here first recapitulate the Jacobi determinant identity applied to the sub-determinants $d_\Gamma$
for unitary matrices \cite{MathWo} which contains much more detailed information than Eq.\
(\ref{coeff_identity}).  As this identity is of great relevance in the paper we also give its proof. We will interpret this identity in terms of periodic orbits
and will discuss the implications
for spectral measures in the remainder of the article. \\
{\bf Theorem}: {\em Let $U$ be a unitary matrix of dimension $N$ with determinant
$\det U = e^{i \phi}$ and $\Gamma
\subset \mathcal{I}\equiv\{1,2,\dots,N\}$ with $n=|\Gamma|$.
Denote the complement
of $\Gamma$ in $\mathcal{I}$ by $\hat\Gamma \equiv \mathcal{I} \setminus \Gamma$.
Then the following identity for the determinants of
the $n\times n$ submatrix $U_\Gamma$ and the $(N-n)\times(N-n)$ submatrix $U_{\hat{\Gamma}} $ holds:
\begin{equation}
  \det U_\Gamma = e^{i \phi} \left(\det U_{\hat\Gamma}\right)^*\ .
  \label{subdet_id}
\end{equation}}
{\em Proof}:
Writing $U$, without loss of generality, in block-form
\begin{equation}
  U=
  \left(
    \begin{array}{cc}
      U_{\Gamma} & V \\
      W & U_{\hat\Gamma}
    \end{array}
  \right)
\end{equation}
the identity can be proven by calculating the determinant of both sides of the matrix identity
\begin{equation}
\left(\begin{array}{cc}
      U_{\Gamma} & V \\
      W & U_{\hat\Gamma}
\end{array}\right)\left(\begin{array}{cc}\mathbbm{1} & W^\dagger \\ 0& U_{\hat\Gamma}^\dagger\end{array}\right)=\left(\begin{array}{cc}U_\Gamma & 0 \\W & \mathbbm{1}\end{array}\right).
\end{equation}
In the last equation the determinant of the first matrix equals $e^{i\phi}$, the second $(\det U_{\hat{\Gamma}})^*$ and the third $\det U_\Gamma$.\hspace*{13.2cm}\rule{0.1in}{0.1in}

This identity implies some fundamental connections
between orbits and pseudo-orbits of dynamical systems, which, in our view, are worth exploring.
We will discuss these implications in the following sections.

As a straightforward consequence, one obtains for the zeta function
\eref{zeta_Ud} for $N$ odd,
\begin{equation}
  \zeta_U(\theta) = \sum_{\Gamma: |\Gamma| \le N/2} \left(
    d_\Gamma e^{-i(\theta+\pi)|\Gamma|} + d_{\Gamma}^* e^{i\phi}
    e^{-i(\theta+\pi)(N-|\Gamma|)}\right)\ .
  \label{zetaU_RS}
\end{equation}
The formula remains true for $N$ even if appropriate care is taken for
contributions with $|\Gamma|=N/2$; only half of these contributions
should be counted and this half needs to be chosen appropriately.
Expression \eref{zetaU_RS} resembles Riemann-Siegel look-alike formulae, see \cite{Berry,Tan91}.

\subsection{Pseudo-orbit expansions in terms of determinants}
\label{sec:pseudo}

In the previous sections, we have expressed the characteristic
polynomial $P_U(z)$ and related expressions in terms of
the determinants $d_\Gamma$. Before we turn
to express the density of states or spectral correlation functions in a
similar fashion, we will consider how the  identity (\ref{subdet_id}) can be
interpreted in a periodic orbit language. To this end, we briefly explain
what we mean by a 'periodic orbit' in terms of a finite matrix and introduce some related
notation. Analogous finite pseudo-orbit expansions in terms of short orbits  have recently been  discussed considering relation (\ref{coeff_identity}) in the context of quantum graphs \cite{BHJ12}. We stress here expansions in terms of sub-determinants which together with Eq.\ (\ref{subdet_id}) give compact expressions for spectral quantities in terms of short periodic orbits.

\subsubsection{Periodic orbit representations}
In the present setting of a unitary $N \times N$ matrix
a periodic orbit $p=\overline{i_1\dots i_n}$
of (topological) length $|p|=n$
is a sequence of $n$ integers $i_m\in \{1,2,\dots, N\}$
where cyclic permutations are identified, e.g. $\overline{134}=\overline{341}$. One should think of a periodic orbit as a set of indices of the matrix $U$
that are visited in a periodic way.
Note that by the term 'periodic orbit', we do not yet refer to classical orbits in the sense of a continuous classical dynamics,  but to products of elements of $U$ with the indices forming a cycle. When considering quantum graphs in Sec.\ \ref{sec:quantum_graphs} these 'periodic orbits' can then indeed be identified with periodic orbits on the graph.
A \emph{primitive} periodic orbit is a sequence $p=\overline{i_1\dots i_n}$ which
is not a repetition of a shorter sequence. If $p$ is not primitive we denote its repetitions
number by $r_p$.
An \emph{irreducible} periodic orbit never returns to the same index, that is,
all $i_m$ are different; the length of an irreducible orbit is at most $N$.
We also define the (quantum) amplitude
\begin{equation}
  t_p= \prod_{m=1}^n U_{i_{m+1} i_m}
\end{equation}
of a periodic orbit. If $p$ is not irreducible one may write its
amplitude as a product of amplitudes of irreducible orbits,
for instance $t_{\overline{1213}}=t_{\overline{12}}t_{\overline{13}}$.

A \emph{pseudo-orbit} $\gamma=p_1^{m_1}p_2^{m_2}\dots p_n^{m_n}$
with non-negative integers $m_l$
is a formal abelian product of periodic orbits $p_l$ with length $|\gamma|=
\sum_l m_l |p_l|$ and amplitude $t_{\gamma}=\prod_l t_{p_l}^{m_l}$. We will say that a
pseudo-orbit is \emph{completely reduced} if it is a formal product
of irreducible orbits and {\em  irreducible} if all $m_l$ are either one or
zero and if any given index appears at most in one $p_l$ with $m_l=1$.

These definitions allow us to write the trace $\mathrm{tr}\ U^n
= \sum_{p:|p|=n} \frac{n}{r_p} t_p
$
as a sum over amplitudes of periodic orbits of length $n$.
Using
$\log \det(\mathbbm{1}- e^{-i\theta}U)= \tr \log (\mathbbm{1}- e^{-i\theta}U)$
in \eref{dos_zeta} and expanding the logarithm
one arrives at the trace formula
\begin{equation}
  \rho(\theta)= \frac{N}{2 \pi} - \frac{1}{\pi}\frac{d}{d\theta}
  \sum_{p \in \mathcal{P}}\sum_{r=1}^\infty \frac{1}{r}t_p^r e^{
    - i |p|\theta}
  \label{trace_formula}
\end{equation}
where sum over $p$ extends over the set of all primitive orbits
denoted by $\mathcal{P}$, and the additional sum is over all repetitions.
Here, like in Eq.\ (\ref{dos_zeta}), it is always understood that $\theta \equiv \theta- i \epsilon$
and the limit $\epsilon \to 0$ is taken.

Performing the sum over repetitions in the trace formula shows that
it is equivalent to an Euler-product type expansion
\begin{equation}
  \zeta_U(\theta)= \prod_{p \in \mathcal{P}} \left(1 -
    t_p e^{-i |p| \theta}\right)\ .
  \label{euler_prod}
\end{equation}
Note that this is an infinite product (which converges for $\epsilon >\ln N$) and analytical continuation is necessary to move back to the
axis $\epsilon = 0$. Such an analytic continuation is of course given by the
expression \eref{zeta_U} which is by definition a finite
polynomial in $z = e^{- i \theta}$. Strong correlations between the amplitudes
of long and short periodic orbits have to exist in order to
reconcile both expressions.
Indeed, large cancellations can be shown to exist by expanding the product (\ref{euler_prod}) and
ordering the terms with increasing orbit length such as in the {\em cycle expansion} proposed in
\cite{Artu}.  After expressing amplitudes of reducible (arbitrarily long) orbits as product of
amplitudes of irreducible (and thus short) orbits, the cancellation mechanism emerges
 \cite{Bog92,Bogomolny}.

Revisiting  Eq.\ (\ref{zeta_Ud}) and observing that each
determinant
$d_\Gamma$ can indeed be written as a sum of $|\Gamma|!$ irreducible pseudo-orbits
$\gamma$ of length $|\gamma|=|\Gamma|$, we obtain
\begin{equation} \label{subdet-pseudo}
  d_\Gamma= \sum_{\gamma \in P_\Gamma} (-1)^{\sigma_\gamma+1} t_{\gamma}\, .
\end{equation}
Here, $P_\Gamma$ is the set of all irreducible pseudo-orbits which cover
the set $\Gamma$ completely, that is, which visit each index in $\Gamma$
exactly once. There is a one-to-one correspondence between
these irreducible pseudo-orbits and permutations. Indeed any permutation
of symbols in $\Gamma$ can be written uniquely as a product of cycles
such that each symbol appears exactly once (up to the ordering of the
cycles which is irrelevant as they commute). Each such product of cycles,
that is, each irreducible pseudo-orbit,
defines a unique permutation. We denote the number of cycles (irreducible
orbits) that make up a given pseudo-orbit $\gamma$ as $\sigma_\gamma$ such that
$(-1)^{\sigma_\gamma+1}$ gives the parity of the permutation.

\subsubsection{Interpretation of the identity (\ref{subdet_id}) in terms of periodic orbits}
Everything said in the previous subsection is valid for general, not necessarily unitary matrices.
Unitarity leads to further non-trivial relations between the amplitudes of short and long orbits such as
the functional equation (\ref{func_eq}) resulting in the relation (\ref{coeff_identity}) for the coefficients of the characteristic polynomial which can in turn be written in terms of orbits.

In Sec.\ \ref{sec:sub_det}, we showed that there is a much more detailed link between sub-determinants and thus between orbits. The identity (\ref{subdet_id}), $d_\Gamma = e^{i \phi} d_{\hat{\Gamma}}^*$, also provides a connection between short and long orbits, but it has in addition an
interesting interpretation in terms of linking
irreducible pseudo-orbits in different parts of 'phase space'. $\Gamma$ and its complement $\hat{\Gamma}$ are by definition disjoint and its union forms the whole set $\mathcal{I}=\{1,2,\dots,N\}$. As stated in Eq.\ (\ref{subdet-pseudo}), $d_\Gamma, d_{\hat{\Gamma}}^*$ consist of all irreducible orbits and pseudo-orbits which completely cover the set $\Gamma$, $\hat{\Gamma}$, respectively, (passing through every index in each of the sets exactly once). The relation \eref{subdet_id} thus implies that the
sum over all irreducible pseudo-orbits that cover
$\Gamma$ is equivalent in weight to the  sum over all irreducible pseudo-orbits that
cover its complement $\hat{\Gamma}$. The two contributions from the pseudo-orbits in $\Gamma$ and the complement $\hat{\Gamma}$ yield together a real term in the spectral determinant, as the contributions from $\Gamma$ and $\hat{\Gamma}$ are complex conjugated to each other up to a global phase. 

The statements up to now refer to unitary quantum maps with the 'periodic orbits' obtained from products of elements of $U$ with indices occurring in  a periodic manner.
Given the close relationship between unitary matrices and quantum maps on the one hand and continuous quantum systems on the other hand, we think that this finding has far reaching consequences.  For continuous dynamics, expressions for spectral quantities in terms of {\it classical} periodic orbits in phase space exist that are asymptotically valid in the limit $\hbar\rightarrow 0$. The relation (\ref{subdet_id}) suggests that the semiclassical weights  associated with periodic orbits and pseudo-orbits of classical maps and flows are spatially correlated at all levels. In particular, summing over all orbits associated with a given subset of the full phase space should yield  a total amplitude which is equal to the contribution from the orbits in the complement and both contributions are phase related. 
In order to make this connection more clear we will consider quantum graphs in Sec.\ \ref{sec:quantum_graphs} where the products of matrix elements of $U$ yield directly expansions in terms of periodic orbits on the graph. Here the periodic-orbit expansions are exact, in contrast to the ones for continuous dynamics.

\subsubsection{Density of states and Newton identities}
We will now consider the density of states and show that it can be
expressed in terms of completely reduced pseudo-orbits.
Equivalently, one can write
it as a sum over products of sub-determinants $d_\Gamma$.
The latter has the advantage that these expressions keep track of
the relation \eref{subdet_id} between individual determinants
which is lost on the level of pseudo-orbit sums.

Making use of Eq.\ (\ref{dos_zeta}), we would like to express $\log\zeta_U$ in terms of sub-determinants. We do this by exploiting the identity
\begin{equation} \label{log-exp}
  - \log (1-x) = \frac{1}{2\pi} \sum_{n=1}^\infty (n-1)! \int_0^{2\pi}
  e^{i \alpha n + xe^{-i \alpha}} d \alpha\, ,
\end{equation}
which formally requires $x<1$.  Note for the derivation of Eq.\ (\ref{log-exp}) that performing the $\alpha$-integral on the right hand side yields the Taylor expansion of the logarithm. Setting $1-x = \sum_\Gamma
  d_\Gamma e^{-i (\theta+\pi) |\Gamma|}$ and using
  (\ref{zeta_Ud}), we formally obtain $\log\zeta_U$ on the left hand side of Eq.\ (\ref{log-exp}).
After expanding out the exponentials, interchanging integration and
summations and carrying out the integration over $\alpha$, one obtains
\begin{equation}
  \log \zeta_U(\theta)=-
  \sum_{\mathbf{m}:|\mathbf{m}|>0}(|{\mathbf{m}}|-1)!
  e^{-i (\theta +\pi)  |\mathbf{m} \Gamma| -i\pi |\mathbf{m}| }
  \prod_{j}
  \frac{d_{\Gamma_j}^{m_j}}{m_j!}\ .
  \label{zeta_U_dGamma}
\end{equation}
Here $\mathbf{m}=(m_1,\dots,m_{2^N-1})$ is a tuple of $2^{N}-1$
non-negative integers and
$\Gamma_1,\dots,\Gamma_{2^N-1}$ is some enumeration of
all non-empty subsets  $\Gamma\subset \mathcal{I}$.
The integer $m_j$ is the multiplicity
of that subset $\Gamma_j$ in one contribution to \eref{zeta_U_dGamma}.
We have also introduced the notations
$|\mathbf{m}|=\sum_{j=1}^{2^N-1} m_j$
and $|\mathbf{m}\Gamma|=\sum_{j=1}^{2^N-1} m_j |\Gamma_j|$.
Note that an analogous equation to \eref{zeta_U_dGamma}
can be given either in a coarser way in
terms of coefficients $a_n$ or in a more detailed way in terms of
products of irreducible pseudo-orbits. The expression in terms of the determinants
$d_\Gamma$ is the most detailed one in which the relation  \eref{subdet_id}
between long and short orbits remains explicit.

Before moving on to the density of states let us consider
the well known expansion
$-\log \zeta_U(\theta) = \sum_{n=1}^{\infty} \frac{1}{n} e^{-i \theta n} \tr U^n$
and compare the coefficients of $e^{- i\theta n}$ with the
corresponding ones in
\eref{zeta_U_dGamma}.
This gives us a direct way to
express
the $n$-th trace in terms of the sub-determinants $d_\Gamma$, that is,
\begin{equation}
  \tr U^n = n (-1)^n \sum_{\mathbf{m}: |\mathbf{m}\Gamma|=n}(|{\mathbf{m}}|-1)! \prod_{j=1}^{2^N-1}
  \frac{(-d_{\Gamma_j})^{m_j}}{m_j!} \ .
  \label{Newton_Gamma}
\end{equation}
This formula is reminiscent of the well-known Newton
identities that express the traces of powers of a square matrix in
terms of the coefficients of the characteristic polynomial, see for example
\cite{Ber84}. Indeed, as mentioned above, there is an expression of the form
\eref{zeta_U_dGamma}
in terms of the coefficients $a_n$ instead of the $d_\Gamma$. The corresponding derivation of the traces leads to the Newton identities. In \eref{Newton_Gamma},
we have  in fact derived a more detailed identity; it allows us to
express the (arbitrarily long) periodic orbits that add up to the
traces $\tr U^n$ explicitly in terms of pseudo-orbits of length smaller than
the matrix size $N$. Furthermore, using \eref{subdet_id},
one has an explicit expression of traces of any power in
terms of pseudo-orbits of maximal length $N/2$.
Ordering the sequence $(\Gamma_1,\dots,
\Gamma_{2^N-1})$ such that it is non-decreasing in length, then
\begin{equation}
  \tr U^n = n (-1)^n \sum_{\mathbf{m}: |\mathbf{m}\Gamma|=n} (|{\mathbf{m}}|-1)!\prod_{j=1}^{2^{N-1}-1}
  \frac{(-d_{\Gamma_j})^{m_j}}{m_j!}
  \prod_{j=2^{N-1}}^{2^{N}-1}
  \frac{(-d_{\hat{\Gamma}_j}^* e^{i \phi})^{m_j}}{m_j!}
  \label{Newton_Gamma2}
\end{equation}
gives the $n$-th trace in terms of contributions which
can be computed from irreducible orbits of length smaller than $N/2$.
In the following it will always be understood that products of the
form appearing in \eref{Newton_Gamma} may be expressed analogously to
\eref{Newton_Gamma2} in terms of short orbits.

Eventually the density of states follows directly from
\eref{zeta_U_dGamma}:
\begin{equation}
  \rho(\theta)= \frac{N}{2\pi} - \mathrm{Im}\!\!\!
  \sum_{\mathbf{m}:|\mathbf{m}|>0}\!\!\! \ (|{\mathbf{m}}|-1)!
  \frac{|{\mathbf m}\Gamma|}{\pi}
  e^{-i (\theta+\pi) |{\mathbf m} \Gamma| -i\pi |\mathbf m| }
  \prod_{j}
  \frac{d_{\Gamma_j}^{m_j}}{m_j!}\ .
  \label{dos_dGamma}
\end{equation}

\section{Spectral fluctuations in terms of sub-determinants and short orbits}
\label{sec:applications}
There is a wide variety of measures for spectral fluctuations which have been considered in the past. We will focus here on expressing spectral measures in terms of sub-determinants and show how the relation \eref{subdet_id} can be used
to understand the contributions of long orbits.
We will in particular consider ensembles of unitary matrices where the ensemble average corresponds to an average
over system parameters or disorder.  In Sec.\ \ref{sec:quantum_graphs}
we will also discuss applications which only
involve a spectral average for a fixed physical system.

\subsection{Spectral fluctuations}
\label{sec:spec_fluct}

For a given ensemble of unitary matrices we denote the ensemble average
of some quantity $f(U)$ by $\langle f(U)\rangle_U$.
In the following, we will consider
cross- or auto-correlation functions for the spectral determinant,
the density of states and other quantities.
We start by giving some general definitions.

\subsubsection{The auto-correlation function of the spectral determinant}

This auto-correlation function has previously been considered from a RMT-perspective in Refs.\ \cite{Fritz_Marek,Kettemann} and semiclassically in diagonal approximation \cite{Kettemann,gregor_auto,Muller} and beyond \cite{Waltner}.
It is defined in terms of $Z(\theta)$ given in (\ref{eq5}) as
\begin{eqnarray} \label{auto}
  A&=& \left\langle \frac{1}{ 2\pi}\int d \theta\ Z_U
    \left(\theta+ \frac{s \pi}{N}\right)
    Z_U\left(\theta- \frac{s \pi}{N}\right)\right\rangle_U\nonumber \\
  &=& e^{i s \pi}\sum_{n=0}^N \left \langle |a_n|^2 \right \rangle_U
  e^{-i \frac{2\pi s n}{N}}\,.
\end{eqnarray}
 In particular, $A(s)$ is the generating function for the variance
$\langle |a_n|^2 \rangle$ of the coefficients of the
characteristic polynomial. Note that $|a_n|^2= |a_{N-n}|^2$ ensures that
$A(s)$ is a real function. In terms of sub-determinants we find
\begin{equation}
  \left\langle | a_n |^2 \right \rangle
  = \sum_{\Gamma, \Gamma':|\Gamma|=|\Gamma'|=n}
  \left\langle d_{\Gamma}d_{\Gamma'}^* \right \rangle_U \ .
  \label{an_variance}
\end{equation}
We will show below that this reduces to the diagonal sum $\Gamma=\Gamma'$
for some specific ensembles.

\subsubsection{Parametric cross correlation for the spectral determinant}

Here we use explicitely the more detailed character of the identity (\ref{subdet_id}) compared to the one given in Eq.\ (\ref{coeff_identity}) by considering a spatially inhomogeneous perturbation acting on {\it different} contributions to the spectral determinant of the {\it same} length in a different manner. Let $U$ be a fixed unitary matrix and define $U_\tau:= e^{i \tau P_v} U$
where $\tau$ is a real parameter
and $P_v$ is the projector onto the $v$-th
basis state; the corresponding matrix is zero everywhere apart from one unit
entry at the $v$-th diagonal position.
Physically one may think of the
parameter $\tau$ as a variation of a local magnetic field.
Denoting the
corresponding coefficients of the characteristic polynomial as
\begin{equation}
  a_n(\tau)= \sum_{\stackrel{\Gamma: |\Gamma|=n,}{v \notin \Gamma}}
  d_\Gamma  + e^{i \tau}
  \sum_{\stackrel{\Gamma: |\Gamma|=n,}{v \in \Gamma}}
  d_\Gamma \, ,
\end{equation}
we consider the following
parametric correlation function for the spectral determinant:
\begin{eqnarray}
  B(\tau) & =& \left\langle \frac{1}{ 2\pi}\int d \theta\ Z_{U(\tau)}
    \left(\theta\right)
    Z_{U}\left(\theta\right)\right\rangle_U\nonumber \\
 & =& e^{- i \tau/2} \sum_{n=0}^N \left \langle a_n(\tau) a_n(0)^*
  \right \rangle_U\ .
  \label{Btau}
\end{eqnarray}
The above expression reduces the problem to the parametric correlations
of the coefficients $a_n(\tau)$ which can be  expressed as
\begin{equation}
  \left \langle a_n(\tau) a_n(0)^*
  \right \rangle_U
  =
  \sum_{\Gamma':|\Gamma'|=n}\left(
  \sum_{\stackrel{\Gamma: |\Gamma|=n,}{v \notin \Gamma}}
  \left \langle d_\Gamma d_{\Gamma'}^* \right\rangle  + e^{i \tau}
  \sum_{\stackrel{\Gamma: |\Gamma|=n,}{v \in \Gamma}}
  \left\langle d_\Gamma d_{\Gamma'}^* \right\rangle \right) \ .
  \label{ataua}
\end{equation}
The two inner sums are here
restricted to sets $\Gamma$ of size
$|\Gamma|=n$ such that
the marked $v$-th basis state is not in $\Gamma$
for the first inner sum and the marked basis state
is an element of $\Gamma$ for the second inner sum.
The outer
sum over $\Gamma'$ is only restricted by $|\Gamma'|=n$.

For the quantity $B(\tau)$ we can now reduce the number of terms to $n \le N/2$ by using the relation
$d_{\Gamma}(\tau)= e^{i(\phi+\tau)} d_{\hat{\Gamma}}^*(\tau)$, Eq.\ \eref{subdet_id}. Note that the relation for the $a_n$  in Eq.\ (\ref{coeff_identity}) would not be sufficient here, as the different components contributing to $a_n$ are exposed to different magnetic fields.


\subsubsection{The spectral two-point correlation function and the form factor}

The spectral two-point correlation function is defined as
\begin{equation}
  R_2(s):= \overline{\Delta}^2 \left\langle
    \frac{1}{2\pi} \int_0^{2\pi} d\theta
    \rho(\theta+ \overline{\Delta} s/2)
    \rho(\theta- \overline{\Delta} s/2) \right\rangle_U -1
\end{equation}
where $\overline{\Delta}= \left(\frac{1}{2\pi}\int_0^{2\pi}
  \rho(\theta) \right)^{-1}= 2\pi /N$
is the mean spacing between eigenphases.
Expanding the density of states in terms of traces
and performing the integral over $\theta$,
one obtains the standard expression
\begin{equation}
  R_2(s)= \frac{2}{N}
  \sum_{n=1}^\infty
  \cos \left( s \frac{2\pi n }{N} \right)
  K_n \, ,
\end{equation}
where
\begin{equation}
  K_n=
  \frac{1}{N}
  \left\langle
    \left| \tr U^n \right|^2
  \right\rangle_U
\end{equation}
is known as the form factor. The form factor played an important role in understanding universal and non-universal aspects of spectral
statistics; here we give a new representation in terms of sub-determinants, that is,
\begin{eqnarray}
  K_n= \frac{n^2}{N}\!\!\!\sum_{\mathbf{m}, \mathbf{m'}: |\mathbf{m}\Gamma|=|\mathbf{m'}\Gamma|=n}\!\!\!
  \left\langle  (|{\mathbf{m}}|-1)!(|{\mathbf{m}}'|-1)!\prod_{j=1}^{2^N-1}
    \frac{(- d_{\Gamma_j})^{m_j} (-d_{\Gamma_j'}^*)^{m_j'}}{m_j! m_j'!}
  \right\rangle_U\ \!\!\!. \nonumber\\
  \label{KnGamma}
\end{eqnarray}
This is an exact expression for the form factor for any
ensemble of unitary matrices. We will show below that for some standard
models, the double sum over multiplicities $\mathbf{m}$ and $\mathbf{m'}$
can be restricted further.

\subsection{Random-matrix theory}

Let us now consider unitary $N \times N$ matrices $U$ which are
distributed according
to the Circular Unitary Ensemble (CUE) -- in other words $U$
has a uniform distribution with respect to the
Haar-measure on the unitary group $U(N)$.
The spectral fluctuations of this ensemble are very well understood
with explicit results for a large number of relevant measures.
These known results have many implications for the statistical properties
of the sub-determinants.

One obtains, for instance, for the correlations of
the coefficients $a_n$ of the characteristic polynomial  \cite{Fritz_Marek}
\begin{equation}
\label{eq100}
  \langle a_n a_{n'}^* \rangle_{\mathrm{CUE}}= \delta_{n n'}
  \qquad  \langle a_n a_{n'} \rangle_{\mathrm{CUE}}=0\, ;
\end{equation}
it is straight forward to extend this result to the correlations between
sub-determinants. Indeed, any average over CUE is necessarily invariant with
respect to conjugation, left multiplication, and right multiplication, that is,
$U \mapsto VUV^\dagger, V U, U V$ with a
unitary matrix $V$. As relation (\ref{eq100}) has to hold also for every transformed $U$, we can choose $V$ diagonal and get
\begin{equation}
  \langle d_\Gamma  d_{\Gamma'}\rangle_{\mathrm{CUE}}=0
\end{equation}
and
\begin{equation}
  \langle d_\Gamma  d_{\Gamma'}^* \rangle_{\mathrm{CUE}}=
  \delta_{\Gamma \Gamma'} c_{\Gamma}\ .
\end{equation}
Note that
\begin{equation}
  \sum_{\Gamma: |\Gamma|=n} c_\Gamma = \langle |a_n|^2 \rangle_{\mathrm{CUE}}=1
\end{equation}
where the sum extends over ${N \choose n}$ contributions.
Moreover, if $\Gamma$ and $ \Gamma'$ have the same size, that is, $|\Gamma|=|\Gamma'|=n$,
invariance of the ensemble average
under conjugation with a permutation matrix implies
\begin{equation}
  c_{\Gamma}=c_{\Gamma'}\equiv c_n \qquad c_n = {N \choose n}^{-1}\ .
\end{equation}
Let us now consider the parametric correlation $B(\tau)$ defined in
\eref{Btau}. Note that it will not depend on the marked basis state, as
 the double sum over $\Gamma$ and $\Gamma'$ in Eq.\ \eref{ataua}
only contains diagonal expressions after the CUE-average. Moreover
among the ${N \choose n}$ subsets $\Gamma$ of a given size $|\Gamma|=n>0$
there are ${ N-1 \choose n-1}$ subsets which contain the marked basis state
and all give the same contribution such that
\begin{equation}
  \left\langle a_n(\tau)a_n^*\right\rangle_{\mathrm{CUE}}=
  \frac{n e^{i \tau}+ N-n}{N}
\end{equation}
and
\begin{equation}
  B(\tau)_{\mathrm{CUE}}= (N+1) \cos (\tau/2) \ .
\end{equation}
Let us finally look at the form factor; the CUE result is
\begin{equation}
  K_{n,\mathrm{CUE}}=
  \begin{cases}
    {n}/{N} & \text{if $n\le N$}\\
    1 & \text{if $n>N$}.
  \end{cases}
  \label{KnCUE}
\end{equation}
We may compare this to the CUE average of the form factor expressed in terms
of sub-determinants \eref{KnGamma}. Invariance of the CUE ensemble
with respect to group multiplication and unitary conjugation restricts the
double sum over multiplicities $\mathbf{m}$ and $\mathbf{m'}$
in \eref{KnGamma}. For example, invariance with respect to multiplication with
diagonal unitary matrices implies that only those pairs can survive, for
which the corresponding product of sub-determinants
$\prod_{j=1}^{2^N-1}d_{\Gamma_j}^{m_j}$
and  $\prod_{j=1}^{2^N-1}d_{\Gamma_j}^{m'_j}$
visit each basis state with the same multiplicity; here, the multiplicity
of a basis state is the number of times a given index appears
in any pseudo-orbit
of the product $\prod_j d_{\Gamma_j}^{m_j}$.
Note that this does not imply $m_j= m_j'$ as there may be many choices
for the multiplicities $m_j$ of the subsets $\Gamma_j$ that
result in the same multiplicities of a basis state.
Comparing the resulting expression with the exact CUE result \eref{KnCUE}
one may obtain a large set of identities that have to be
obeyed by the correlations
among the sub-determinants.

\subsection{Quantum graphs}
\label{sec:quantum_graphs}
\subsubsection{Star graphs - an introduction}

A quantum graph is a model for a quantum particle that is confined to a
metric graph. To keep the discussion simple we will only discuss
star graphs which consist of one central vertex and $N$ peripheral vertices.
Each peripheral vertex is connected to the center by a bond (or edge)
of finite length $0<L_b< \infty$. By $L=\mathrm{diag}(L_1,\dots,L_N)$ we
denote the diagonal matrix that contains the lengths on its diagonal.
On a given bond $b$ we denote by $x_b\in (0,L_b)$ the distance from
the central vertex. A scalar wave function on the graph is a collection of
$N$ complex (square-integrable) functions
$\boldsymbol{\Psi}(x)=(\psi_1(x_1),\dots, \psi_N(x_N))$.
The wave function is required to solve the free stationary Schr\"odinger
equation on each bond, at given energy $E=k^2$. This implies
$\psi_b(x_b)= a_b \left(e^{ikx_b} +  e^{-ikx_b+ 2ikL_b}\right) $
where $a_b$ is the amplitude of the outgoing wave from the central vertex and
we have imposed Neumann
boundary conditions at the peripheral vertices (at $x_b=L_b$).
The matching conditions at the
central vertex are given in terms of a unitary $N \times N$
scattering matrix $S$ which relates the
amplitudes $a_b$ of outgoing waves to the amplitudes $a_b e^{2ikL_b}$
of incoming waves by
$a_b= \sum_{b'} S_{bb'}e^{2ikL_{b'}} a_{b'}$.
Equivalently
\begin{equation}
  \mathbf{a}= U(k) \mathbf{a}
  \label{qg_spectrum}
\end{equation}
for the quantum map
\begin{equation}
  U(k)= T(k) S  \qquad \textrm{where}\quad T(k)=e^{2ikL}\ .
  \label{graph_qmap}
\end{equation}
This implies that the evolution resulting from $U(k)$ consists of scattering events at the vertices and free evolution on the connecting bonds. This propagation can thus be described by the paths on the graph and the spectral quantities can be expressed in terms of sums over pseudo orbits. In contrast to systems with continuous dynamics, the spectral quantities describing graphs possess {\it exact} expressions in terms of periodic orbits. This can be understood by following our derivation of expressions for spectral quantities in terms of $U(k)$.
The condition \eref{qg_spectrum} is only satisfied at discrete values
of the wave number which form the (wave number) spectrum of the graph.
As a side remark let us also note that the above defined quantum map
for a star graph also
describes the quantum evolution on directed graphs with first-order
(Dirac-type) wave operators
and bond lengths $2L_b$ \cite{gregor_auto}. A more general quantum graph
requires a description in terms of a $2N \times 2N$ matrix \cite{Kottos}.

Spectra of quantum graphs and spectra of the associated
unitary quantum maps $U(k)$ have formed a paradigm of quantum chaos
due to the conceptual simplicity of the models. In fact, both types of spectra are
to a large extent equivalent \cite{berkolaiko} and we will focus the present discussion
on the spectrum of the quantum map $U(k)$. It can be considered as an
ensemble of unitary matrices parametrised by $k$. The corresponding average
will be denoted by
\begin{equation}
  \left\langle F\left( U(k) \right) \right\rangle_k=
  \lim_{K\to \infty} \frac{1}{K} \int_0^K dk  F\left( U(k) \right)\ .
\end{equation}
Note that the wave number $k$ enters the quantum map $U(k)=T(k) S$
only through the diagonal factor $T(k)=e^{2iLk}$.

The sets $\Gamma\subset \mathcal{I}$ in this model are one-to-one
related to sub-graphs spanned by the corresponding bonds.
The sub-determinants
$d_\Gamma$ of $U(k)$ can thus be written as
\begin{equation}
  d_\Gamma= e^{i k L_\Gamma} \tilde{d}_\Gamma
\end{equation}
where $L_\Gamma= 2 \sum_{b \in \Gamma} L_b$ is twice the metric length
of the sub-graph connected to $\Gamma$ and $\tilde{d}_\Gamma=
\det S_\Gamma$ is the corresponding sub-determinant of the scattering
matrix $S$.
A generic choice of lengths $L_b$  implies that the lengths are rationally
independent (incommensurate), which will be assumed in the following.
Incommensurability implies that
$\left\langle e^{ik\sum_{b=1}^N m_b L_b  }\right\rangle_k$
does vanish except for $m_b=0$ for all $b=1,\dots,N$.

\subsubsection{Results for general star graphs}
It is straight forward to implement the averages for the
spectral fluctuation measures introduced in Sec.\ \ref{sec:spec_fluct}.
Let us start with the variance
of the coefficients of the characteristic polynomial, Eq.\ \eref{an_variance}, which
build up the auto-correlation function $A(s)$. Due to the difference in the metric lengths
of the corresponding sub-graphs only diagonal
entries in the double sum of Eq.\ \eref{an_variance} survive the average, that is,
\begin{equation}
  \left\langle | a_n |^2 \right \rangle_k
  = \sum_{\Gamma:|\Gamma|=n}
  \left|\tilde{d}_{\Gamma}\right|^2 \ .
  \label{an_variance_qg}
\end{equation}
Note that the expression can not reduce further due to averaging. Contributions
from different sets $\Gamma$ contain orbits of different length, so
non-diagonal contributions made up of products of orbits from different sub-graphs
$\Gamma$ do not survive the average; orbits and pseudo-orbits contained
in $\tilde{d}_\Gamma$ cover the same sub-graph $\Gamma$, and thus have all
the same lengths \cite{Tanner}. In full analogy,
we find
\begin{equation}
  \left \langle a_n(\tau) a_n(0)^*
  \right \rangle_k
  =
  \sum_{\stackrel{\Gamma: |\Gamma|=n,}{v \notin \Gamma}}
  \left| \tilde{d}_\Gamma \right|^2  + e^{i \tau}
  \sum_{\stackrel{\Gamma: |\Gamma|=n,}{v \in \Gamma}}
  \left| \tilde{d}_\Gamma \right|^2
  \label{ataua_qg}
\end{equation}
for the parametric correlations \eref{ataua}.
In contrast to the CUE result this will generally depend
on the marked $v$-th basis state.

Furthermore, for the spectral two-point correlations, the form factor reduces to
\begin{eqnarray}
  K_n= \frac{n^2}{N} \sum_{L \in \mathcal{L}_n}
  \sum_{\stackrel{\mathbf{m}, \mathbf{m'}: }{L_{\mathbf{m}\Gamma}=L_{\mathbf{m'}\Gamma}=L}}(|{\mathbf{m}}|-1)!(|{\mathbf{m'}}|-1)!
  \prod_{j=1}^{2^N -1}
  \frac{(-\tilde{d}_{\Gamma_j})^{m_j}(-\tilde{d}_{\Gamma_j}^*)^{m_j'}
  }{m_j! m_j'!}\nonumber\\
  \label{Kn_qg}
\end{eqnarray}
where the $\mathcal{L}_n$ is the set of all lengths that are a sum
of $n$ (not necessarily different) bond lengths of the graph.
We have used the short-hand notation $L_{m\Gamma}=\sum_{j=1}^N m_j L_{\Gamma_j}$.
Note that equality of metric length $L_{\mathbf{m}\Gamma}=L_{\mathbf{m'}\Gamma}$ implies
equality of the topological length
$|\mathbf{m}\Gamma|=|\mathbf{m'}\Gamma|$ while the opposite is not true.
Eq.\ \eref{Kn_qg} expresses the form factor as a sum over
all possible metric lengths with a fixed number $n$ of bonds
and a sum over pairs of completely reduced pseudo-orbits
of topological length $n$
of the same metric length.

\begin{figure}
\vspace{2cm}
\begin{center}
\vspace*{-2cm}
\includegraphics[width=8cm]{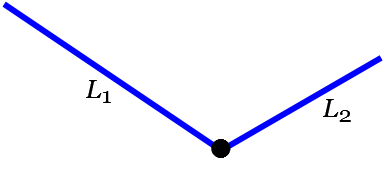}
\caption{The two-star graph consists of one vertex and two bonds labeled 1 and 2.}
\label{2graph}
\end{center}
\end{figure}

\subsubsection{The two-star graph}
It is instructive to work out
the simplest non-trivial case $N=2$ in more detail.
In this case the only choices for $\Gamma$ are the empty set,
$\Gamma_1=\{1\}$, $\Gamma_2=\{2\}$ and $\Gamma_3=\{1,2\}$ with
lengths $L_{\Gamma_1}=2L_1$, $L_{\Gamma_2}=2L_2$, and
$L_{\Gamma_3}=2(L_1+L_2)$, see Fig.\ \ref{2graph}. The zeta function can be described in terms of
the sub-determinant $\tilde{d}_{\Gamma_1}=  S_{11}$ which
is just the reflection amplitude from the first bond and the determinant $\tilde{d}_{\Gamma_3}= \det S= e^{i
\phi}$ alone; without loss of generality we set $\det S=1$.
The remaining relevant sub-determinant is
given by $\tilde{d}_{\Gamma_2}= S_{22}=\tilde{d}_{\Gamma_1}^*$ due to \eref{subdet_id} and $\det S=1$.

Let us consider how an expansion of the zeta function in terms of periodic orbits and pseudo-orbits as discussed in Sec.\ \ref{sec:pseudo} would look like. By expanding the product (\ref{euler_prod}) and reordering the terms according to a  cycle expansion \cite{Artu}, one obtains, for example,
\begin{eqnarray}
\label{cyc-exp}
  \zeta_{U(k)}(\theta) &=&1 - (t_1 + t_2)  e^{-i\theta} - (t_{12} - t_1 t_2) e^{-2 i \theta} \\
  &-& (t_{112} - t_1 t_{12} + t_{122} - t_{12} t_2) e^{-3 i \theta} - \ldots. \nonumber
  \end{eqnarray}
Writing this out in terms of determinants yields instead
\begin{eqnarray}
\label{eq1000}
  \zeta_{U(k)}(\theta)
  &=&1 - e^{i 2L_1k-i\theta}\tilde{d}_{\{1\}}-
  e^{i 2L_{2} k-i\theta}\tilde{d}_{\{1\}}^*
  +e^{i 2(L_1+L_2) k - 2 i \theta} \\
  &=& 2 e^{i(L_1+L_2)k - i \theta}\left(\cos(2k(L_1+L_2) - \theta) -
  \Re \left( e^{i (L_1 - L_2)k} \tilde{d}_{\{1\}}\right) \right) \,\nonumber .
\end{eqnarray}

The cancellation of contributions from longer pseudo-orbits $|p| > 2$ appearing in the expansion, Eq.\ (\ref{cyc-exp}), becomes obvious when writing periodic orbits as  completely reduced pseudo-orbits. For example, the contribution $t_{122}$ from the orbit $\{122\}$ is exactly canceled by $t_{12}t_2$ from the pseudo-orbit $\{12\}\{2\}$ contributing just with opposite sign. By applying this cancellation mechanism recursively, i.e.\ reducing the orbits step by step, also the cancellation of contributions from longer pseudo-orbits can be understood. A similar cancellation argument is also used by the cycle expansion.
This is however different for the contributions $t_{12} - t_1t_2$ in (\ref{cyc-exp}). In this case a reduction of the connected orbit leading to cancellation is not possible.

The equivalence between pseudo-orbits on a subset $\Gamma$ and its complements can be made more explicit. The
first and the last term in Eq.\ (\ref{eq1000}) resulting from pseudo-orbits of zero length and the length of
the full graph, respectively, both have modulus of order 1 and yield a real contribution to
$\zeta_{U(k)}(\theta)$ when the phase factor $e^{i\left(L_1+L_2\right)k-i\theta}$ is taken out. The same
holds for the second and the third contributions to Eq.\ (\ref{eq1000}) from the orbits on the set
$\Gamma= \{1\}$ and $\Gamma=\{2\}$,
respectively. Here, the identity (\ref{subdet_id}) comes in to yield a real contribution (up to an overall pre-factor).

For this simple example, we can calculate the spectral measures discussed in Sec.\ \ref{sec:spec_fluct} explicitly. For the auto-correlation function, Eq.\ (\ref{auto}), one obtains
\begin{equation}
  A(s)= 2\cos(\pi s)+ 2|\tilde{d}_{\Gamma_1}|^2\ .
\end{equation}
For the parametric correlation function, Eq.\ (\ref{Btau}), we consider $U(k;\tau)
= \mathrm{diag}(e^{i\tau},1) U(k)$. One  then obtains
\begin{equation}
  B(\tau)= 2\left(1+|\tilde{d}_{\Gamma_1}|^2 \right)\cos( \tau/2) \ .
\end{equation}
Eventually, let us consider the form factor $K_n$
for a given $n$ as presented in (\ref{Kn_qg}). It contains a sum
over pairs of multiplicities
$\mathbf{m}=(m_1,m_2,m_3)$ and $\mathbf{m'}=(m_1',m_2',m_3')$.
Both sums are restricted to have the same topological length
$|\mathbf{m}\Gamma|=|\mathbf{m'}\Gamma|=n$ which implies
two restrictions, namely
$m_1+m_2+2m_3=n=m_1'+m_2'+2m_3'$. Furthermore only pairs
of multiplicities contribute that have the same metric length
$L_{\mathbf{m} \Gamma}=L_{\mathbf{m'} \Gamma}$ or
$L_1(m_1-m_1'+m_3-m_3')+L_2(m_2-m_2'+m_3-m_3')=0$.
The latter implies $m_1+m_3=m_1'+m_3'$ and
$m_2+m_3=m_2'+m_3'$. Only three of these four restrictions on pairs
of orbits are independent. The form factor can then be written as
\begin{eqnarray}
\label{form3}
     K_n&=&\frac{n^2}{2}
    \sum_{0\le m_3 \le n/2 \atop 0\le m_3'\le n/2}
    \sum_{ 0\le m_2 \le n-2m_3\atop 0\le m_2'\le n-2m_3'} \delta_{m_2+m_3,m_2'+m_3'}(n-m_3-1)!(n-m_3'-1)!\nonumber
    \\[0.3cm]
    &&\times
    \quad {n-2m_3 \choose m_2}
    {n-2m_3' \choose m_2'}
    \frac{
      (-1)^{m_3+m_3'}|\tilde{d}_{\Gamma_1}|^{2(n-m_3-m_3')}}{
      m_3! m_3'! (n-2m_3)!(n-2m_3')!} \, .
\end{eqnarray}
Writing the Kronecker as
\[ \delta_{m_2+m_3,m_2'+m_3'}= \frac{1}{2\pi}
\int_0^{2\pi} d\alpha\, e^{i \alpha(m_2-m_2'+m_3-m_3')}\]
 makes it possible to sum
over $m_2$ and $m_2'$ independently. With
\[\sum_{m_2=0}^{n-2m_3}e^{i \alpha (m_2+m_3)}{n-2m_3 \choose m_2}=
(2 \cos(\alpha/2))^{n-2m_3}e^{i \alpha n/2}\]
and \[\sum_{m_2'=0}^{n-2m_3'}e^{-i \alpha (m_2'+m_3')}{n-2m_3' \choose m_2'}=
(2 \cos(\alpha/2))^{n-2m_3'}e^{-i \alpha n/2},\] we obtain
\begin{eqnarray}
\label{form4}
 K_n&=& \frac{n^2}{2}\int_0^{2\pi}d\alpha\!\!\!
  \sum_{0\le m_3 \le n/2 \atop 0\le m_3'\le n/2}\!\!\!
  {n-m_3 \choose m_3} {n-m_3' \choose m_3'}\frac{(-1)^{m_3+m_3'}|{\tilde d}_{\Gamma_1}|^{2(n-m_3-m_3')}}{\left(n-m_3\right)\left(n-m_3'\right)}\nonumber\\
    &&\times
  \left(2\cos\frac{\alpha}{2}\right)^{2(n-m_3-m_3')} .
\end{eqnarray}
The sums with respect to $m_3$ and $m_3'$ can be performed by using \cite{Prud}
\begin{eqnarray}
\sum_{0\le m_3 \le n/2}\frac{(-1)^{m_3}}{\left(n-m_3\right)}{n-m_3 \choose m_3}x^{2m_3}&=&\frac{1}{2^nn}\left[\left(1+\sqrt{1-4x^2}\right)^n+\right.\nonumber\\&& \left.\left(1-\sqrt{1-4x^2}\right)^n\right].
\end{eqnarray}
This yields for $K_n$
\begin{eqnarray}
\label{form1}
K_n&=&\frac{1}{4\pi}\int_0^{2\pi}d\alpha\left[\left(\cos\frac{\alpha}{2}\left|\tilde{d}_{\Gamma_1}\right|+
\sqrt{\cos^2\frac{\alpha}{2}\left|\tilde{d}_{\Gamma_1}\right|^2-1}\right)^{2n}+\right.\nonumber\\ &&
\left.\left(\cos\frac{\alpha}{2}\left|\tilde{d}_{\Gamma_1}\right|-
\sqrt{\cos^2\frac{\alpha}{2}\left|\tilde{d}_{\Gamma_1}\right|^2-1}\right)^{2n}+2\right].
\end{eqnarray}
The constant term at the end describes the behavior for $n\gg 1$, the other two contributions describe oscillations around the asymptotic value $K_n = 1$. By taking into account that the arguments of the square roots above are negative, we can rewrite the expression as
\begin{equation}
\label{form2}
K_n=1+\frac{1}{2\pi}\int_0^{2\pi} d\alpha\cos \left[2n\arccos
\left( \left|\tilde{d}_{\Gamma_1}\right|\cos \frac{\alpha}{2}\right)\right].
\end{equation}
The expression in Eq.\ (\ref{form2}), which we obtained from periodic-orbit expansions, coincides for all $\tilde{d}_{\Gamma_1}$ with the result obtained  in \cite{SchanzI} starting from the eigenvalues of the quantum scattering map. This is the first derivation of the result (\ref{form2}) from periodic-orbit expressions for general $\tilde{d}_{\Gamma_1}$. In \cite{SchanzI}, a derivation based on periodic-orbit expressions was only done for $\left|\tilde{d}_{\Gamma_1}\right|=1/\sqrt{2}$.

Note that in contrast to \cite{Braun_Haake_RS_maps} we also take into account contributions beyond the diagonal
approximation. Due to the factor $(-1)^{m_3+m_3'}$ appearing in the Eqs.\ (\ref{form3}, \ref{form4}) the ones with $m_3-m_3'$ odd contribute with negative signs leading together with the ones with $m_3-m_3'$ even to a form factor smaller than expected in diagonal approximation; as expected it tends to $K =1$  for $n \rightarrow \infty$.

\section{Conclusions}

The goal of this article is two-fold: first of all, we advocate considering sub-determinant
expansions for spectral functions and statistical measures such the density of states or various
correlation functions. This makes it possible to separate out contributions which vanish after
averaging and those whose non-diagonal contributions survive averaging. Secondly, we considered a
sub-determinant identity due to unitarity which makes it possible to give much more detailed relations between short
and long orbits on a graph than considered before. In particular, this identity implies that contributions to
the characteristic polynomial originating from irreducible pseudo-orbits of a certain sub-graph have the same weight
as the irreducible pseudo-orbits of the complement of that sub-graph and are additionally linked through a common phase factor. Previously, only relations between the overall contributions from pseudo orbits of a certain length and the complementary length were studied.
The identity leads to simplified expressions for the characteristic polynomial, the Newton identities and the
spectral density. Furthermore we study the effect of this identity on spectral correlation functions such as
the auto-correlation function of the characteristic polynomial, the parametric cross correlation function
and the spectral form factor.

We derive explicit expressions using sub-determinant expansions for a simple model, star graphs
consisting of $N$ bonds connected by a single vertex. We then work out in more detail the simplest
case $N=2$. The
identity (\ref{subdet_id}) is essential to obtaining the behaviour of correlation functions for small energy differences
or large times. It captures additional correlations between orbits of different length and needs
to be taken into account when singling out correlated orbits which survive averaging. This is especially important when spatial inhomogeneities affect {different} parts of the phase space in {different} ways.

In this context several potentially interesting extensions arise: taking the semiclassical limit on both
sides of Eq.\ \eref{Newton_Gamma2}, the two expressions are semiclassically not obviously identical. The left-hand
side leads to the Gutzwiller trace formula which contains orbits of arbitrary length while the right-hand side
contains pseudo-orbits of finite length (and their repetitions). For short orbits $n \ll N$, one may argue
that the two expressions have a semiclassically small difference, for longer orbits this is far less obvious.

A second point concerns the exponential proliferation of the number of orbits in the standard trace formulae.
It is tamed to a certain degree when using sub-determinants by the fact that different contributions contribute
with different signs. Thus the sub-determinant expressions contain large fluctuations. Understanding overall
cancellations is an interesting task. For example, the form factor for the two-star graph for large $n$
contains positive and negative contributions which on their own grow as $n \to \infty$ while their
difference remains $\mathcal{O}(1)$ as can be checked from the expressions for $K_n$ given above.

The analysis of spectral correlations focused here on the general unitary case. It would be interesting
to include the effect of self crossings of orbits that allow for partners traversing parts of
the diagram in different directions. This would capture effects arising due to time reversal symmetry.

For graphs the supersymmetry technique gives an alternative approach to obtain universal results \cite{SGAA}.
With supersymmetry one may derive universality under sufficiently nice conditions, however, rigorous
proofs are still not available. The main obstacle for the supersymmetric approach seems to be repetitions
which are difficult to incorporate correctly \cite{GKP}. The proposed approach may help us
understanding the effect repetitions on spectral correlation functions.

\section{Acknowledgments}

{
We thank Christopher Eltschka and Jack Kuipers for useful discussions.
DW and KR thank the {\em Deutsche Forschungsgemeinschaft} (within research unit FOR 760) and DW the {\em Minerva foundation}
for financial support. SG acknowledges support by the {\em EPSRC research network} 'Analysis on Graphs' (EP/I038217/1).
}


\end{document}